\documentclass[aps, pra, reprint, superscriptaddress]{revtex4-2}

\usepackage{graphicx}
\usepackage{dcolumn}
\usepackage{bm}
\usepackage{upgreek}
\usepackage{color,soul}
\usepackage{amsmath}

\begin{document}

\preprint{}

\title{A magneto-ionic synapse for reservoir computing}

\author{Sreeveni Das}
\affiliation{NanoSpin, Department of Applied Physics, Aalto University School of Science, P.O. Box 15100, FI-00076 Aalto, Finland}
\author{Rhodri Mansell}
\affiliation{NanoSpin, Department of Applied Physics, Aalto University School of Science, P.O. Box 15100, FI-00076 Aalto, Finland}
\email[Corresponding author: ] {rhodri.mansell@aalto.fi}
\author{Luk{\'{a}}{\v{s}} Flaj{\v{s}}man}
\affiliation{NanoSpin, Department of Applied Physics, Aalto University School of Science, P.O. Box 15100, FI-00076 Aalto, Finland}
\author{Maria-Andromachi Syskaki}
\affiliation{Singulus Technologies AG, 63796 Kahl am Main, Germany}
\author{J\"{u}rgen Langer}
\affiliation{Singulus Technologies AG, 63796 Kahl am Main, Germany}
\author{Sebastiaan van Dijken}
\email[Corresponding author: ]{sebastiaan.van.dijken@aalto.fi}
\affiliation{NanoSpin, Department of Applied Physics, Aalto University School of Science, P.O. Box 15100, FI-00076 Aalto, Finland}

\begin{abstract}
Neuromorphic computing aims to revolutionize large-scale data processing by developing efficient methods and devices inspired by neural networks. Among these, the control of magnetism through ion migration has emerged as a promising approach due to the inherent memory and nonlinearity of ionically conducting and magnetic materials. In this work, we present a lithium-ion-based magneto-ionic device that uses applied voltages to control the magnetic domain state of a perpendicularly magnetized ferromagnetic layer. This behavior emulates the analog and non-volatile properties of biological synapses and enables the creation of a simple reservoir computing system. To illustrate its capabilities, the device is used in a waveform classification task, where the voltage amplitude range and magnetic bias field are tuned to optimize the recognition accuracy.
\end{abstract}

\maketitle

\section{Introduction}
Neuromorphic computing, inspired by the architecture and functionalities of the human brain, aims to create systems capable of efficient information processing, learning, and memory tasks \cite{Roy2019,Markovic2020,Yang2020,Zhu2020,Schuman2022}. A defining feature of neuromorphic architectures is the integration of processing and memory operations within the same physical layer, overcoming the limitations imposed by their separation in conventional semiconductor-based computers. This multidisciplinary field encompasses diverse approaches, including systems based on synaptic \cite{Kuzum2012,Yang2013,Wang2017,Majumdar2019,Cao2021,Yu2021} and neuronal \cite{Torrejon2017,Han2022} analogs, probabilistic computing \cite{Misra2023}, and reservoir computing \cite{LUKOSEVICIUS2009127,Tanaka2019}. In synaptic-based systems, the core functionality relies on the synaptic weight, which quantifies the strength of connections between nodes. Adjusting this weight through potentiation or depression enables synapses to process and store data within a network \cite{Roy2019,Markovic2020,Yang2020}. Reservoir computing has gained attention for its simplicity and efficiency, requiring training only the readout layer. It utilizes the system's inherent nonlinear dynamics and short-term memory to perform computational tasks \cite{LUKOSEVICIUS2009127,Tanaka2019}. 

Magnetic materials are of significant interest for neuromorphic computing applications \cite{Grollier2020} due to their intrinsic memory properties, nonlinear magnetization dynamics, and stochastic behavior. Among these, magnetic films hosting skyrmions, topologically protected spin textures, have emerged as versatile candidates for emulating synaptic \cite{Huang2017,Song2020} and neuronal \cite{Li2017,Chen2018} behavior, as well as enabling probabilistic \cite{Pinna2018,ZazvorkaSkyrmionDiffusion} and reservoir \cite{Prychynenko2018,Bourianoff2018,Pinna2020,Raab2022,Yokouchi2022,Sun2023} computing devices. In these systems, skyrmions are manipulated using external stimuli, such as magnetic fields \cite{Yokouchi2022}, electric currents \cite{Song2020,Prychynenko2018,Bourianoff2018,Pinna2020,Raab2022}, or applied voltages \cite{Sun2023}. Skyrmion-based artificial synapses and reservoir computing devices often encode synaptic weights and reservoir outputs in skyrmion density \cite{Song2020,Yokouchi2022,Sun2023}, providing a discrete and scalable metric well-suited for neuromorphic applications. These magnetic devices exploit the nonlinear dynamics of skyrmion nucleation and annihilation, combined with their intrinsic memory retention, to perform computational tasks. Outputs are typically read through magneto-optical imaging or anomalous Hall effect measurements, enabling effective integration into neuromorphic architectures.   

\begin{figure}[htbp]
    \centering
    \includegraphics[width=1.0\linewidth]{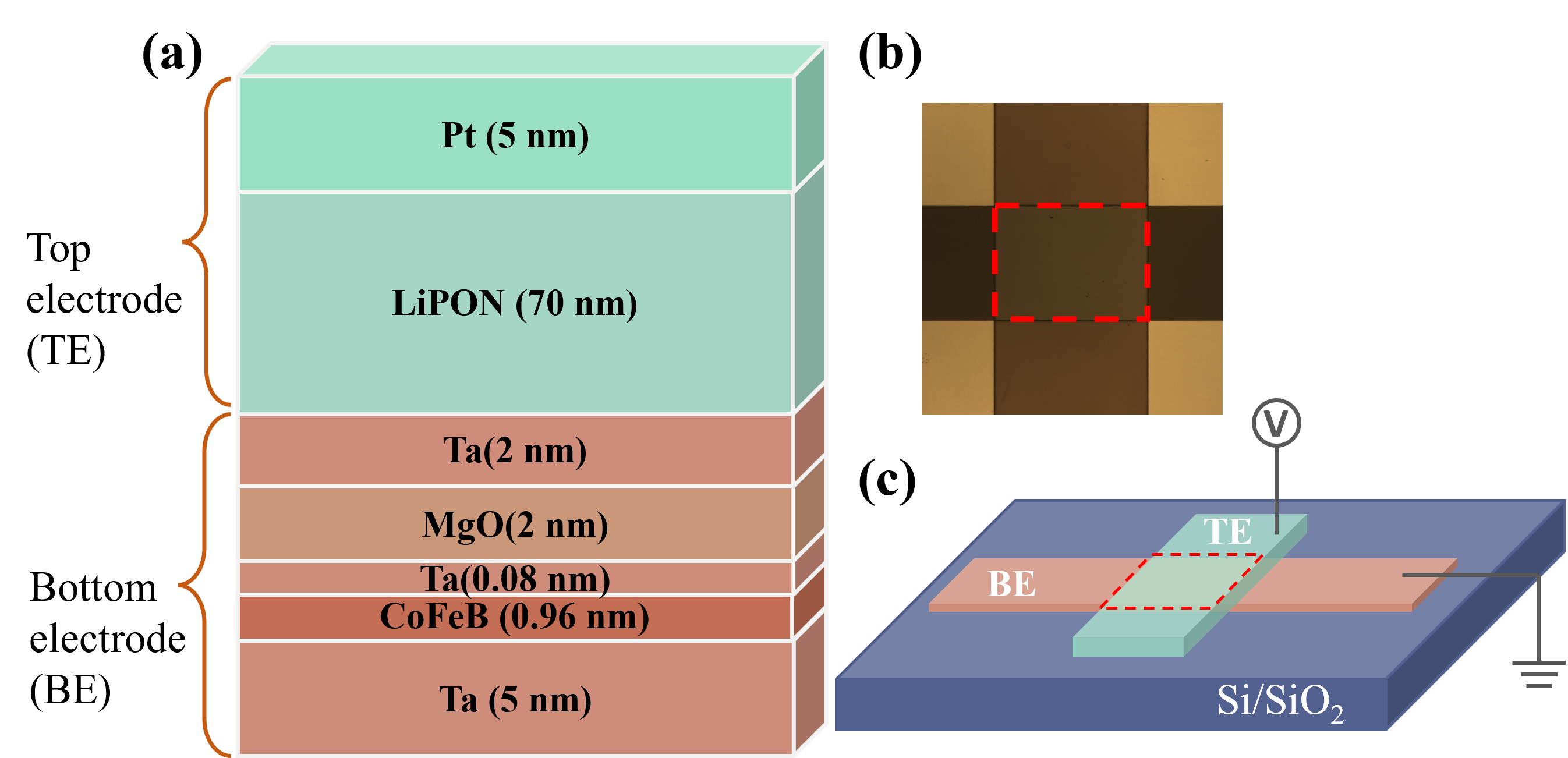}
    \caption{(a) Cross-sectional schematic of the multilayer magneto-ionic device operating as a solid-state supercapacitor. (b) Optical microscopy image of the device, with the junction area (105\(\times\)80 \(\upmu\)m\(^2\)) marked in red. (c) Schematic of the crossbar junction design. The bottom electrode (BE) consists of Ta (5 nm) / Co\(_{20}\)Fe\(_{60}\)B\(_{20}\) (0.96 nm) / Ta (0.08 nm) / MgO (2 nm) / Ta (2 nm), while the top electrode (TE) includes LiPON (70 nm) / Pt (5 nm). When a positive voltage is applied, Li$^+$ ions migrate from the LiPON layer into the bottom electrode, enabling modulation of the magnetic domain state in the CoFeB layer.}
    \label{fig1}
\end{figure}

\begin{figure*}[hbtp]
    \centering
    \includegraphics[width=1.0\linewidth]{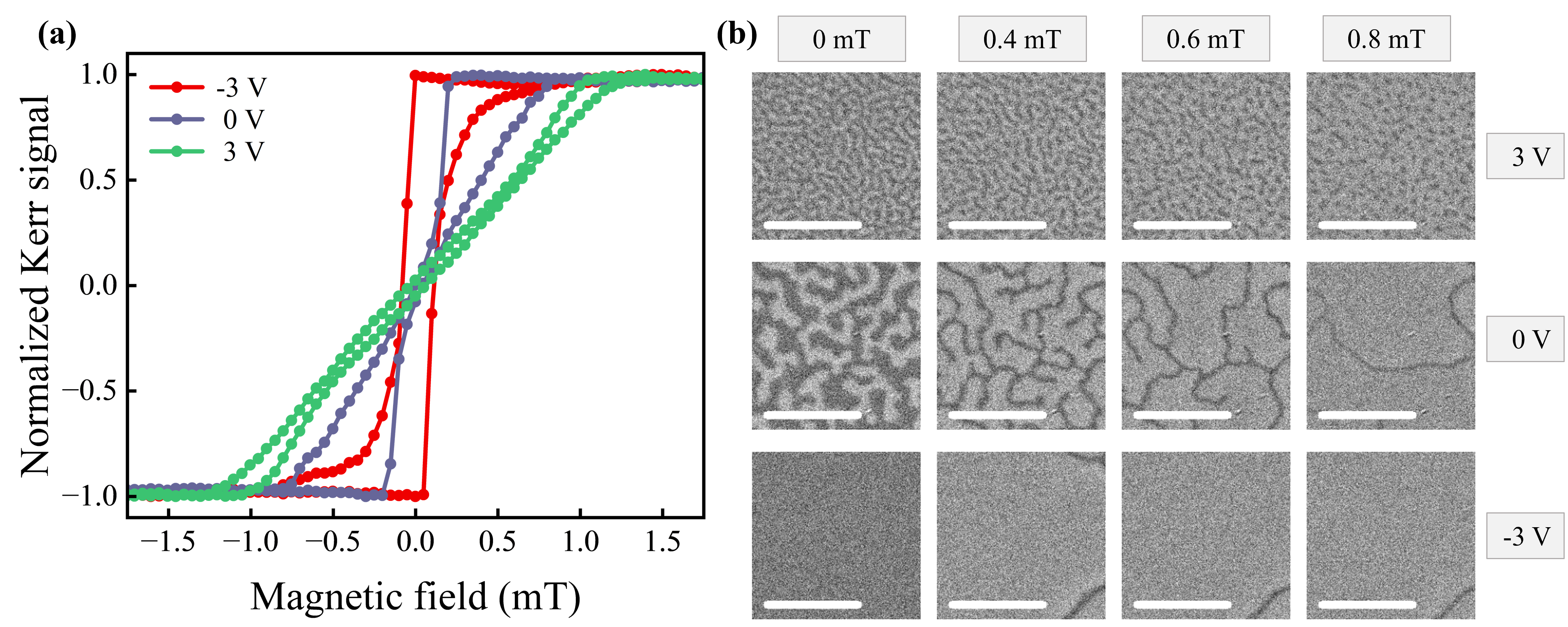}
    \caption{(a) Polar MOKE hysteresis loops measured on the magneto-ionic device under applied voltages -3 V, 0 V and 3 V. (b) MOKE images showing the evolution of magnetic domains for various combinations of voltage and magnetic field. The images were captured as the magnetic field was swept from negative to positive values. The scale bar represents 10 \(\upmu\)m.}
    \label{fig2}
\end{figure*}

Magneto-ionics, which utilizes voltage-driven ion migration to manipulate magnetism in gated structures, offers an energy-efficient mechanism compatible with conventional electronics. The combined use of magnetic fields and voltages to control magnetization or magnetic textures in magneto-ionic systems presents new opportunities for neuromorphic computing \cite{Mishra2019,Monalisha2023,Monalisha2024}, including task-adaptation through memory and nonlinearity tuning \cite{Lee2024}. Ionic modulation of key magnetic properties, such as saturation magnetization, magnetic anisotropy, and exchange coupling, has been demonstrated with various ionic species, including oxygen \cite{Bi2014,Bauer2015}, hydrogen \cite{Tan2018,Huang2021}, nitrogen \cite{deRojas2020,Rojas2022}, and lithium \cite{Ameziane2022,Ameziane2023a}. Additionally, magneto-ionics has been employed to induce skyrmion nucleation and annihilation with the application of small voltages \cite{Fillion2022, Ameziane2023}. 

In this paper, we present a magneto-ionic synapse that utilizes voltage-driven Li$^+$ ion migration in a solid-state supercapacitor structure to reversibly manipulate the domain state of a perpendicularly magnetized CoFeB layer. By applying controlled voltages and magnetic fields, we modulate the density of stripe domains and magnetic skyrmions, effectively mimicking synaptic weight adjustments during potentiation and depression. Through a tailored sequence of voltage pulses, we achieve precise control over the domain and skyrmion nucleation and annihilation dynamics, enabling fine-tuned synaptic weight modulation with varying memory capacity. The nonlinear response and intrinsic memory of the magneto-ionic system make it suitable for reservoir computing applications \cite{Liu2022}. To demonstrate this, we perform a sequential waveform classification task \cite{Pinna2020,Yokouchi2022,Sun2023,Zhong2021} and analyze the dependence of recognition accuracy on voltage amplitude range and magnetic bias field. The device achieves 100$\%$ recognition accuracy for current waveforms and approximately 70$\%$ for previous waveforms while operating in a mixed skyrmion-stripe domain state. Our results demonstrate the potential of solid-state magneto-ionic structures to emulate synaptic behavior and perform reservoir computing tasks, with performance optimized through precise control of voltage- and field-driven physical processes.  

\section{Results}
Figure\ \ref{fig1}(a) shows a cross-sectional schematic of the multilayer magneto-ionic device. The stack consists of a Ta (5 nm) / Co\(_{20}\)Fe\(_{60}\)B\(_{20}\) (0.96 nm) / Ta (0.08 nm) / MgO (2 nm) / Ta (2 nm) bottom electrode, paired with a lithium phosphorus oxynitride (LiPON) (70 nm) / Pt (5 nm) top electrode. All layers were grown via magnetron sputtering at room temperature. The device was patterned into a crossbar junction through a two-step lithography process. First, the bottom stack was grown onto a Si/SiO\(_2\) substrate and patterned into 80-\(\upmu\)m-wide electrodes using Ar ion beam milling. Subsequently, the LiPON/Pt layers were sputtered and lifted off to form 105-\(\upmu\)m-wide top electrodes. Additional fabrication details are provided in Appendix A: Methods. 

\begin{figure*}[tbp]
    \centering
    \includegraphics[width=0.95\linewidth]{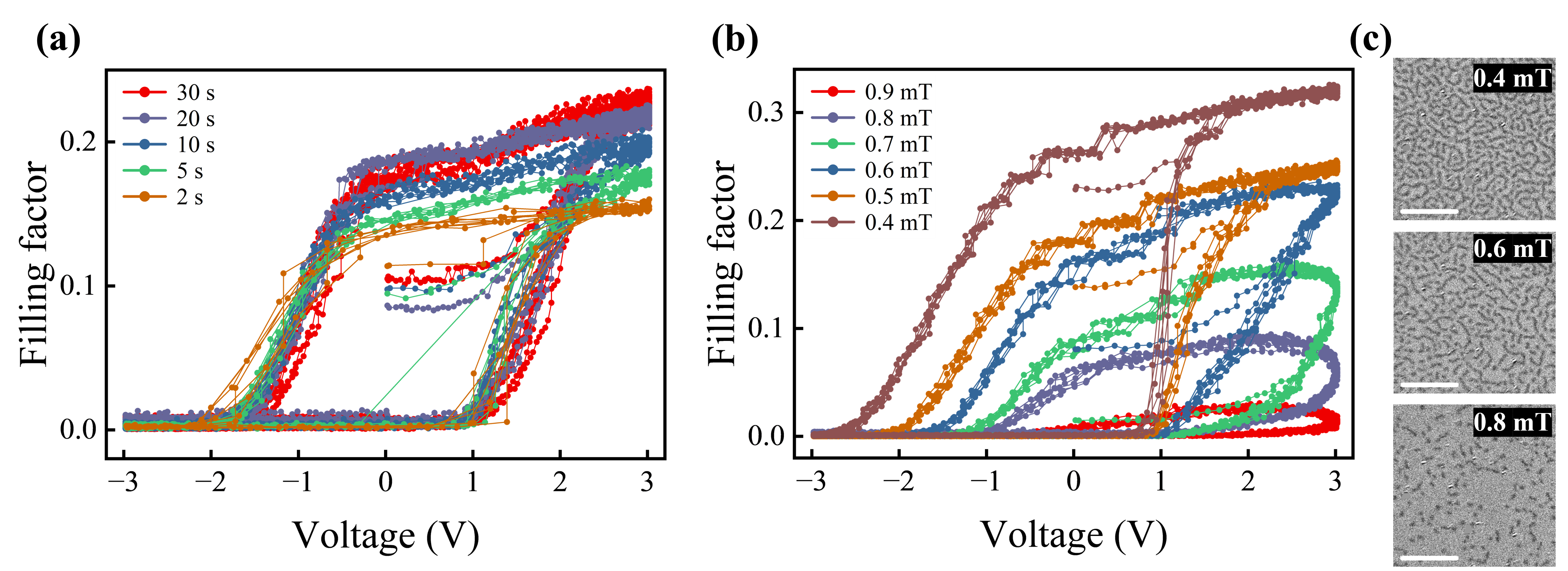}
    \caption{(a) Filling factor of the magneto-ionic device during voltage cycling under a magnetic field of 0.6 mT. The voltage was varied sinusoidally with distinct oscillations periods as indicated in the legend. (b) Filling factor of the device during voltage cycling at a fixed oscillation period of 10 s under different applied magnetic fields. (c) MOKE microscopy images showing the magnetic microstructure during the backward voltage sweep at 2 V for distinct magnetic fields. The scale bar represents 10 \(\upmu\)m.}
    \label{fig3}
\end{figure*}

Figure\ \ref{fig1}(b) shows an optical microscopy image of the crossbar device, with the junction area highlighted in red. Figure\ \ref{fig1}(c) presents a schematic of the junction, including its electrical contact configuration. In experiments, the voltage is applied to the top electrode while the bottom electrode is grounded. The LiPON layer acts as a solid-state electrolyte, facilitating reversible Li$^+$ ion migration. Under a positive voltage, Li$^+$ ions migrate from the LiPON layer into the bottom electrode; a negative voltage reverses the migration. Cyclic voltammetry (CV) measurements (\textcolor{blue}{Supplementary Fig.\ S1}) show open loops without redox peaks, consistent with previous reports on similar magneto-ionic devices \cite{Ameziane2023}. These results confirm that the device operates as supercapacitor \cite{Gogotsi2018}, where an electric double layer forms, and Li$^+$ ions migrate into the bottom electrode under positive voltage. Operating as a supercapacitor, rather than a battery, enables faster voltage-driven modulation of magnetic properties \cite{Monalisha2024,Ameziane2023,Molinari2017}. This work analyzes and discusses data from two similar crossbar junctions, each measuring 105\(\times\)80 \(\upmu\)m\(^2\). 

Figure\ \ref{fig2}(a) displays magnetic hysteresis loops of the magneto-ionic device recorded using magneto-optic Kerr effect (MOKE) microscopy, under varying out-of-plane magnetic fields and applied bias voltages. Corresponding MOKE images, illustrating the evolution of magnetic domain states under different voltage and field conditions, are shown in Fig.\ \ref{fig2}(b). At a positive voltage of 3 V, Li\(^+\) ions migrate into the bottom electrode, reducing the perpendicular magnetic anisotropy (PMA) of the CoFeB layer \cite{Ameziane2022,Ameziane2023}. This reduction leads to a slanted hysteresis loop. As the magnetic field increases from 0 mT to 0.8 mT, dense stripe domains transition into a skyrmion state. At 0 V, reducing the magnetic field from saturation causes a sudden transition into a stripe domain state before the field reaches zero. Reversing the magnetic field decreases the stripe domain density, but skyrmion formation is not observed. Compared to the 3 V condition, stripe domains at 0 V are less dense. Under a negative voltage of -3 V, Li\(^+\) ions are driven out of the bottom electrode, increasing the PMA of the CoFeB layer. In this state, the CoFeB magnetization switches abruptly at low magnetic fields, resulting in a square-shaped hysteresis loop. The observed variations in polar MOKE hysteresis loops and magnetic domain structures arise from the reversible modulation of PMA driven by Li\(^+\) ion migration \cite{Ameziane2022,Ameziane2023}. Lower PMA under positive voltages reduces the energy of magnetic domain walls, making the formation of stripe domains and skyrmions energetically more favorable \cite{Schott2017,Mansell2023}.

Figure\ \ref{fig3}(a) illustrates the magnetic response of the system to a sinusoidal input voltage  V(\(t\)) = \(A \mathrm{sin}(\frac{2\pi}{T} t)\), where the amplitude \(A\) was fixed at 3 V and the oscillation period \(T\) was varied. To quantify this response, MOKE microscopy images were captured at a rate of $12-16$ frames per second, and the magnetic \textit{filling factor} was extracted for each frame. This filling factor, representing the spatial occupancy of magnetic features such as stripe domains and skyrmions, was determined by applying a binary mask to the MOKE images (see Appendix A: Methods for details). All measurements in Fig.\ \ref{fig3}(a) were conducted under a 0.6 mT magnetic field. The hysteretic behavior of the filling factor during voltage sweeps highlights the system's full reversibility and intrinsic memory. As the voltage increases from -3 V, inverse magnetic domains begin to nucleate at approximately 1 V. Their density and size grow progressively with increasing voltage. Upon reversing the sweep direction after reaching 3 V, these domains gradually shrink and are annihilated completely between -1.5 V and -2.0 V. The sweep rate significantly affects the magnetic domain evolution. Faster sweep rates yield fewer and smaller domains during the forward sweep and require larger negative voltages to reach magnetic saturation during the reverse sweep. Since Li\(^+\)-ion-induced PMA modulation occurs on sub-second timescales \cite{Ameziane2023}, the observed dependence on sweep rate is attributed to thermally activated magnetic processes, including magnetic domain nucleation, annihilation, and domain wall motion.  

\begin{figure*}[tbp]
    \centering
    \includegraphics[width=1.0\linewidth]{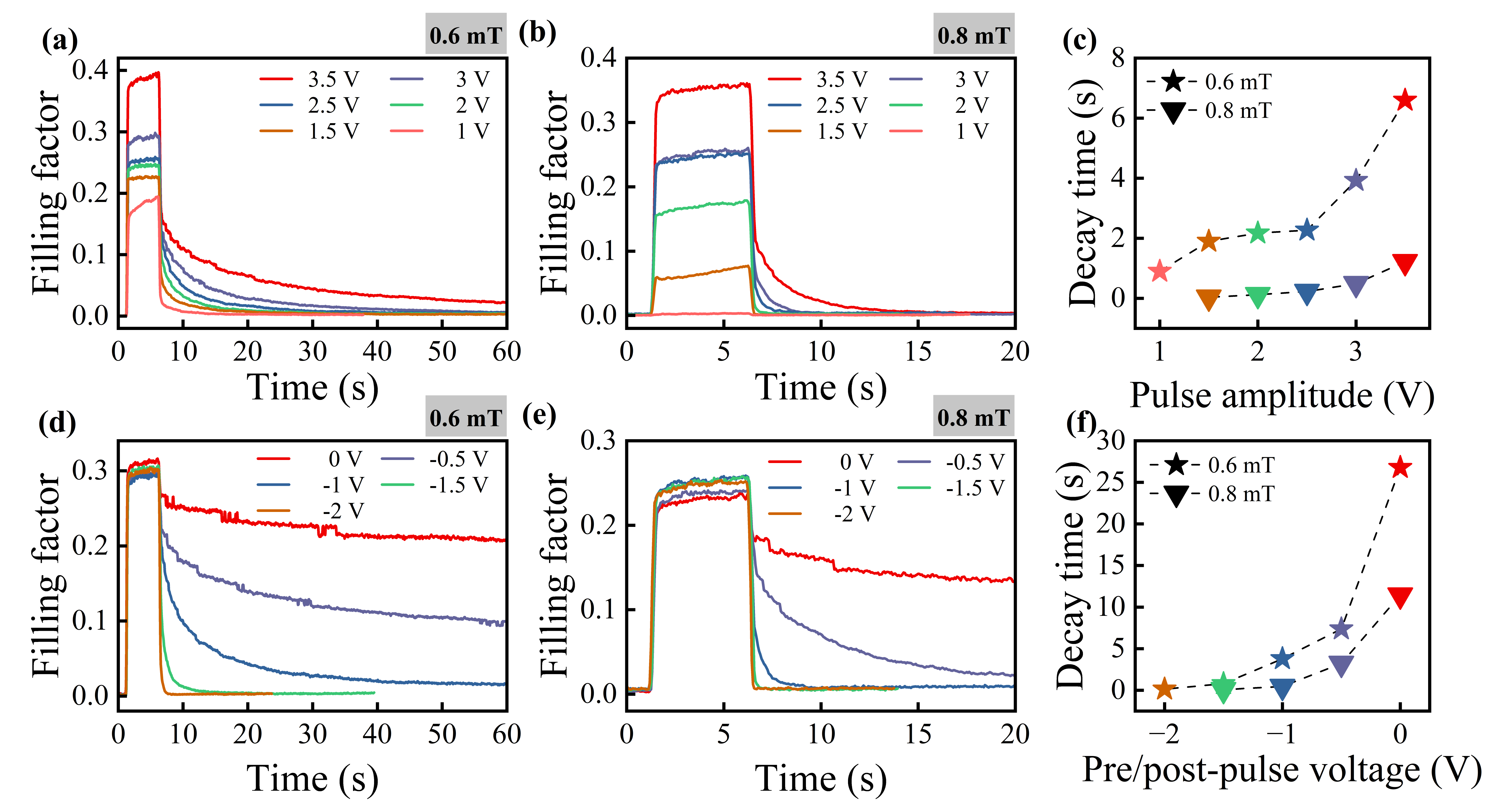}
    \caption{(a, b) Filling factor response to a 5 s positive voltage pulse measured at (a) 0.6 mT and (b) 0.8 mT magnetic field. The pulse amplitude varies between 1 V and 3.5 V, with the pre- and post-pulse voltage fixed at -1 V. (c) Decay time constant of the filling factor after the pulse, extracted by fitting an exponential decay to the data in (a) and (b). (d, e) Filling factor response to a 5 s positive voltage pulse measured at (d) 0.6 mT and (e) 0.8 mT magnetic field. Here, the pulse amplitude is fixed at 3 V, while the pre- and post-pulse voltage is varied between -2 V and 0 V. (f) Decay time constant of the filling factor after the pulse, derived from the data in (d) and (e).}
    \label{fig4}
\end{figure*}

Figure\ \ref{fig3}(b) explores the influence of the applied magnetic field on the filling factor during voltage cycling, with a fixed oscillation period of $T=10$ s. At 0.4 mT, stripe domains nucleate rapidly during the forward voltage sweep, starting at around 1 V, resulting in a sharp increase in the filling factor. In contrast, at 0.9 mT, only a few skyrmions form above 2 V, keeping the filling factor relatively low. Between these two extremes, a gradual transition is observed at positive voltage, with high-density stripe domains at low fields evolving into low-density skyrmions at high fields. This behavior is visualized through MOKE images in Fig.\ \ref{fig3}(c). During the reverse sweep, skyrmions are more readily annihilated than stripe domains. For instance, at 0.9 mT, skyrmions vanish at approximately -0.5 V, whereas stripe domains persist until about -2.5 V at 0.4 mT. Notably, some stripe domains at low magnetic fields transform into skyrmions before annihilation, as detailed in \textcolor{blue}{Supplementary Fig.\ S2}. The data in Fig.\ \ref{fig3} reveal that the filling factor of the magneto-ionic device can adopt multiple discrete values depending on the applied voltage, sweep rate, and magnetic field. This tunability underscores its potential as a metric for emulating synaptic behavior. 
 
Next, we investigate the response of the magneto-ionic device to single voltage pulses. Figure\ \ref{fig4}(a) shows the evolution of the filling factor under a 5 s positive voltage pulse at 0.6 mT magnetic field. Before the pulse, a -1 V bias initializes the system in a fully saturated magnetization state, which is restored after the pulse. When the pulse is applied, the filling factor increases sharply due to rapid nucleation of stripe domains and skyrmions (see \textcolor{blue}{Supplementary Fig.\ S3} for MOKE images). This initial surge is followed by a gradual rise throughout the pulse duration. The system's response, which becomes more pronounced with higher pulse amplitudes, resembles synaptic potentiation. This behavior originates from a reduction in PMA in the CoFeB layer as Li\(^+\) ions migrate into the bottom electrode, lowering the energy barrier for domain and skyrmion nucleation. When the voltage returns to -1 V, the filling factor decreases in two distinct phases: an initial sharp drop, followed by a slower decay. The timescale of this decay depends on the pulse amplitude, with higher amplitudes resulting in longer decay times. 

At a higher magnetic field of 0.8 mT (Fig.\ \ref{fig4}(b)), domain and skyrmion nucleation is suppressed, particularly at lower pulse amplitudes, while the post-pulse decay accelerates. Figure\ \ref{fig4}(c) quantifies this behavior by presenting decay time constants derived from exponential fits to the data recorded at 0.6 mT and 0.8 mT fields. The decay time increases approximately exponentially with pulse amplitude for both fields, with higher fields consistently showing faster decay. MOKE images (\textcolor{blue}{Supplementary Figs.\ S4 and S5}) reveal that stripe domains annihilate or transform into skyrmions immediately after the pulse, leaving the slower decay phase dominated by skyrmion annihilation. This two-phase decay reflects a marked difference in the energy barriers for annihilating stripe domains versus skyrmions, offering a means to tailor the device's nonlinearity and memory capacity.    

Figure\ \ref{fig4}(d-f) shows results from a similar experiment where the pulse amplitude is fixed at 3 V, but the pre- and post-pulse voltage levels are varied between -2 V and 0 V. During the pulse, the filling factor reaches comparable levels regardless of the initial voltage, with slightly higher values observed at 0.6 mT (Fig.\ \ref{fig4}(d)) compared to 0.8 mT (Fig.\ \ref{fig4}(e)). However, the post-pulse decay rate strongly depends on the post-pulse voltage, as summarized in Fig.\ \ref{fig4}(f). Overall, the data in Fig.\ \ref{fig4} highlight the versatility of the magneto-ionic device, enabling precise control over the temporal dynamics of the filling factor through voltage and magnetic field. This flexibility positions the device as a promising candidate for neuromorphic computing applications, where tunable nonlinearity and intrinsic memory are crucial.

\begin{figure}[tbp]
    \centering
    \includegraphics[width=1\linewidth]{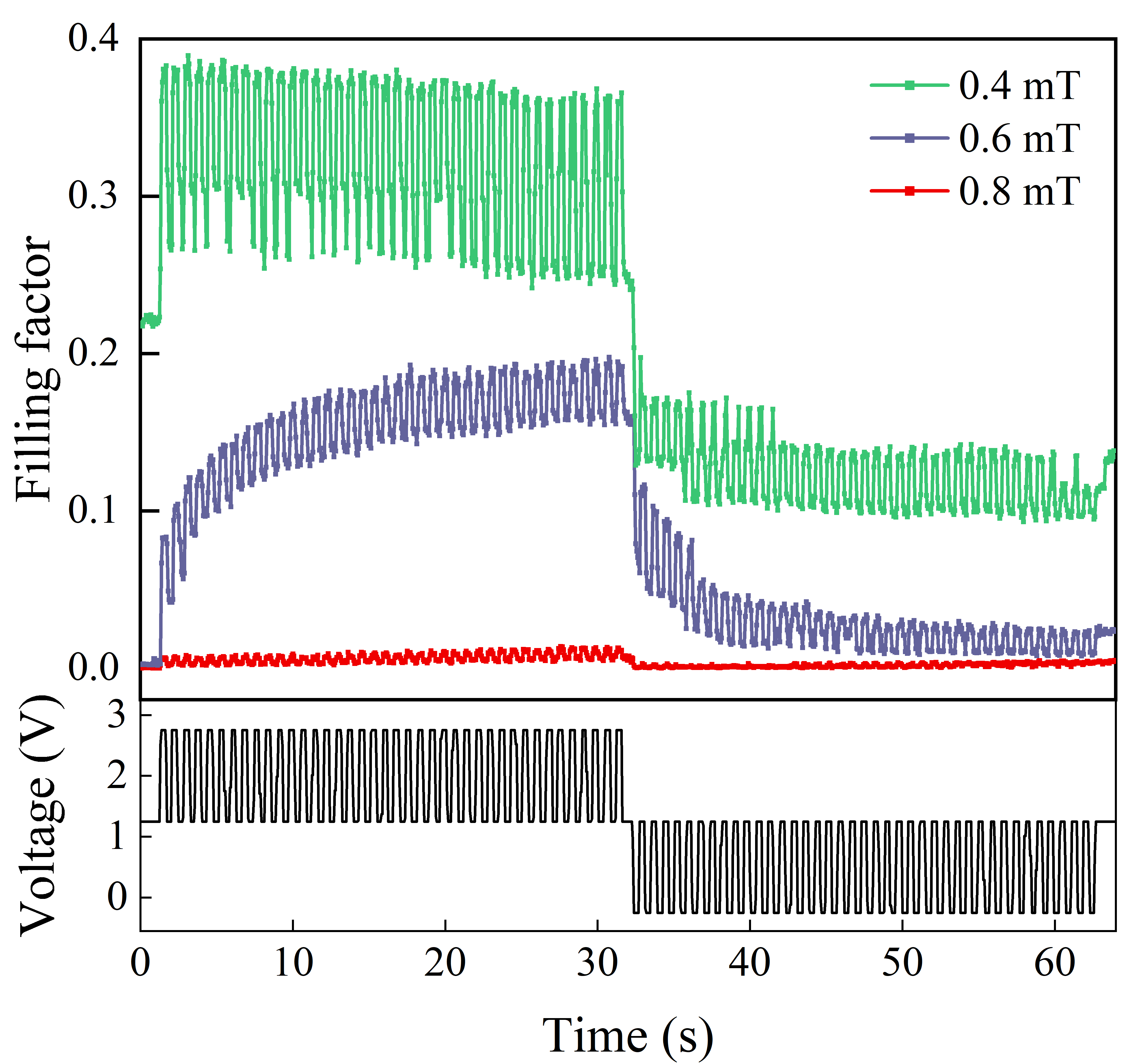}
    \caption{Filling factor response of the magneto-ionic device to a sequence of voltage pulses. Potentiation is induced using 40 pulses alternating between 1.25 V and 2.75 V, followed by depression triggered by 40 pulses oscillating between 1.25 V to -0.25 V. Each pulse has a duration of 375 ms. Results are shown for applied magnetic fields of 0.4 mT, 0.6 mT, and 0.8 mT.}
    \label{fig5}
\end{figure} 

\begin{figure*}[tbp]
    \centering
    \includegraphics[width=0.95\linewidth]{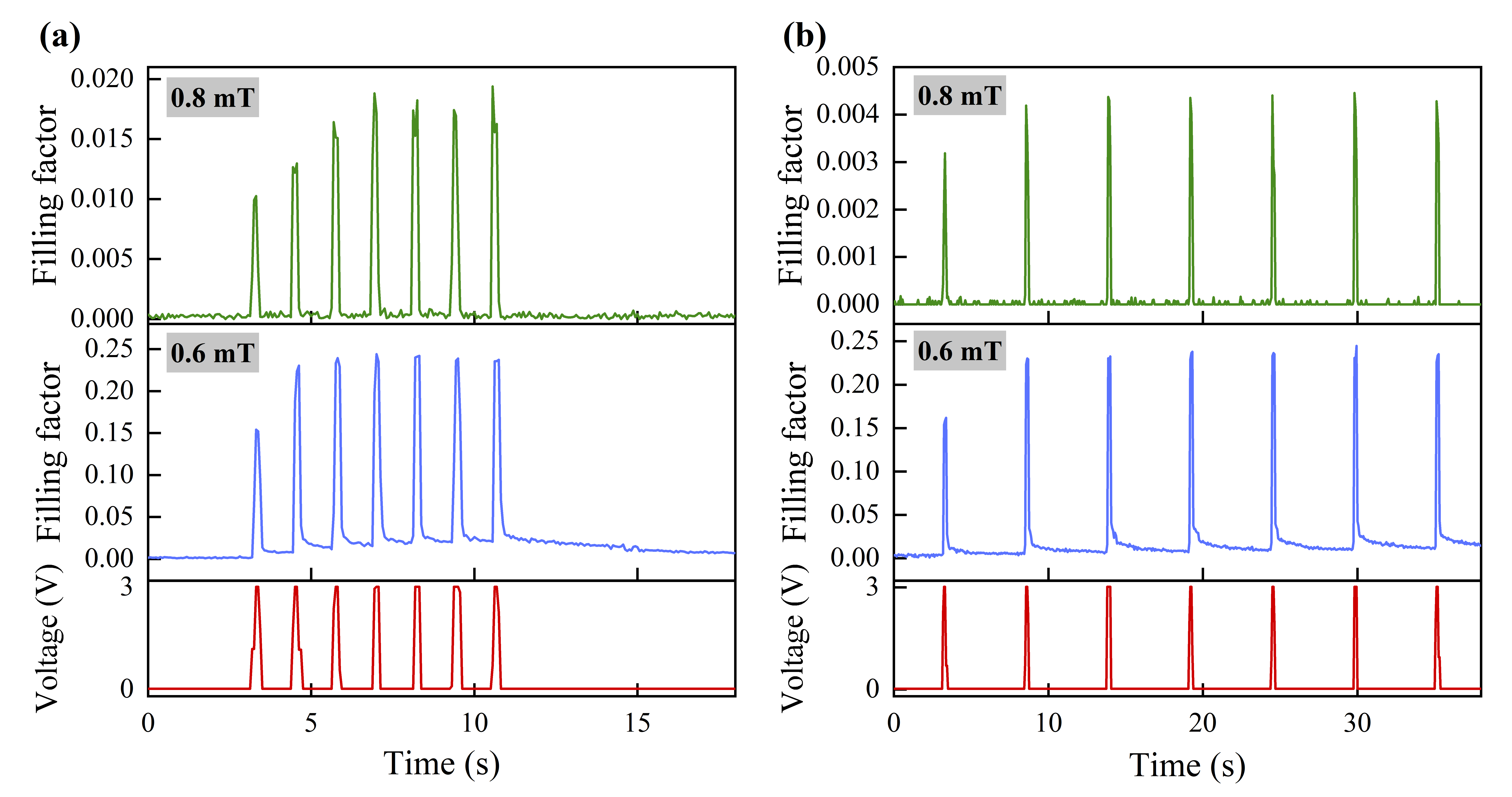}
    \caption{(a, b) Filling factor response of the magneto-ionic device to a series of 3 V voltage pulses under magnetic fields of 0.6 mT and 0.8 mT. Each pulse has a duration of 200 ms, with intervals of 1 s in (a) and 5 s in (b).}
    \label{fig6}
\end{figure*} 

In biomimetic synapses, synaptic weights are often modulated via spike-timing-dependent plasticity using a series of activation pulses \cite{Yu2021}. Figure\ \ref{fig5} illustrates how a sequence of voltage pulses influences the filling factor of the magneto-ionic device under varying magnetic fields. The experiments consist of 80 consecutive pulses, each lasting 375 ms. During potentiation (40 pulses), the voltage alternates between 1.25 V and 2.75 V. This is followed by a depression sequence (40 pulses) with the voltage oscillating between 1.25 V and -0.25 V. The response of the device to these voltage pulses is highly sensitive to the applied magnetic field. At 0.4 mT, where stripe domains dominate, the filling factor undergoes significant modulation during both potentiation and depression. Notably, the first pulse in each sequence triggers a pronounced response, setting the stage for subsequent changes. At 0.6 mT, where stripe domains coexist with skyrmions, the behavior becomes more gradual and controlled, with smaller oscillation amplitudes compared to 0.4 mT. In contrast, at 0.8 mT, where skyrmions dominate the magnetic structure, the filling factor exhibits minimal variation in response to the pulse sequence. This reduced effect arises from the higher energy barrier associated with skyrmion nucleation, which limits changes to the saturated magnetization state at the applied pulse amplitudes. 

Figure\ \ref{fig6} explores the device's response to a shorter sequence of 7 voltage pulses (amplitude: 3 V, duration: 200 ms) with pulse intervals of 1 s (Fig.\ \ref{fig6}(a)) and 5 s (Fig.\ \ref{fig6}(b)), under magnetic fields of 0.6 mT and 0.8 mT. At 0.6 mT, a 1 s interval results in incomplete recovery of the filling factor between pulses, causing it to progressively increase, a behavior reminiscent to long-term potentiation of the synaptic weight. Extending the pulse interval to 5 s eliminates this cumulative effect, allowing the filling factor to reset after each pulse. In both cases, the sharp increase in the filling factor after each pulse, indicative of short-term potentiation, is consistent across pulses except the first. Switching to a skyrmion-dominated state at 0.8 mT alters the dynamics. Here, the filling factor increases less with each pulse, and the 1 s interval allows for a full reset between pulses due to the low energy barrier for skyrmion annihilation. Interestingly, the peak filling factor continues to grow with successive pulses, particularly at shorter intervals, despite full resets between pulses. This behavior can be attributed to the dynamics of Li\(^+\) ion migration under applied voltage. During high-frequency pulsing (short intervals), Li\(^+\) ions migrate rapidly into the bottom electrode at 3 V but return more slowly to the LiPON layer at 0 V. This imbalance leads to ion accumulation in the bottom electrode, progressively reducing the PMA and saturating the process over time. The magnetic energy landscape, shaped by the applied magnetic field and PMA, dictates the filling factor response. At 0.8 mT, where skyrmion nucleation is constrained by a high energy barrier, the gradual reduction in PMA facilitates enhanced skyrmion formation with successive pulses. In contrast, at 0.6 mT, stripe domains emerge rapidly and reach a dense configuration after only two pulses, despite the relatively higher PMA. This interplay between ion migration dynamics and the distinct energetics of stripe domains and skyrmions allows the device to finely tune its short-term and long-term potentiation and depression behaviors. 

The structure and properties of the magneto-ionic device make it attractive for integration into synaptic device architectures. To achieve this, the current method of detecting the synaptic weight via magneto-optic imaging must be replaced by a faster and more scalable electrical readout mechanism. The junction geometry of the device is particularly well-suited for forming crossbar networks, akin to those employed in neuromorphic memristor networks \cite{Yang2013,Han2022} and magnetic random access memories (MRAMs) \cite{Jung2022}. In the present study, the filling factor serves as a proxy for the average magnetization of the CoFeB layer. This parameter could be directly and rapidly measured through the tunneling magnetoresistance (TMR) effect within a magnetic tunnel junction. Such a configuration would not only enable high-speed electrical readout but also facilitate the integration of magneto-ionic devices into large-scale, energy-efficient neuromorphic computing systems. 

\section{Reservoir computing}
\begin{figure*}[htbp]
    \centering
    \includegraphics[width=1.0\linewidth]{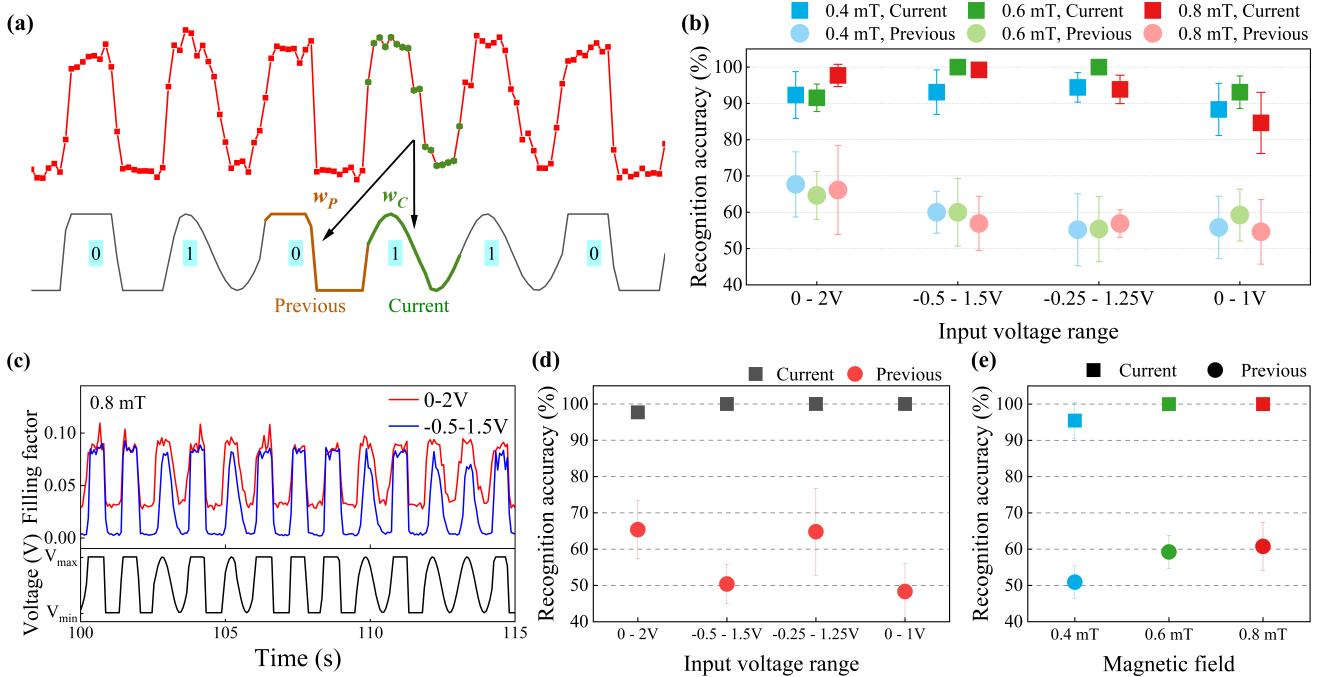}
    \caption{(a) Illustration of the input voltage signal consisting of randomly ordered sine and square waveforms (bottom) and the corresponding filling factor output signal (top). Highlighted sections in green and orange represent examples of current and previous waveforms used in the classification task. Green markers on the output signal indicate sampling points determined by the MOKE microscopy frame rate. The classification task involves weights (\textit{w}$_{C}$ and \textit{w}$_{P}$) for the current and previous waveforms, with binary indicators (1 for sine, 0 for square) shown in blue boxes. (b) Waveform recognition accuracy for various input voltage ranges and magnetic bias fields. (c) Filling factor response to randomly ordered sine and square waveforms for input voltage ranges of $0-2$ V and -$0.5-1.5$ V under a magnetic field of 0.8 mT. (d) Recognition accuracy using combined training across multiple magnetic fields (0.4 mT, 0.6 mT, and 0.8 mT) for each voltage range, demonstrating significant improvements in performance. (e) Recognition accuracy achieved using combined training across all four voltage ranges for each magnetic bias fields, emphasizing the benefit of multi-range training.}
    \label{fig7}
\end{figure*}

Access to multiple magnetic states and precise control over their temporal dynamics via voltage-pulse sequences and magnetic bias fields make the magneto-ionic device a promising platform for reservoir computing. By nonlinearly transforming input voltages into a high-dimensional representation within the magnetic state of the CoFeB layer, which exhibits short-term memory, the device enables computation. To demonstrate its computational capability, a sine-square waveform classification task was performed \cite{Pinna2020,Yokouchi2022,Sun2023,Zhong2021}, where sine and square voltage waveforms were applied to the device in random order. Figure\ \ref{fig7}(a) illustrates the task: a random voltage waveform (small selection shown in the bottom panel) was used as input, and the measured device response (filling factor, top panel) was used to classify the current (green) and previous (orange) voltage waveforms as either sine or square. The dataset comprised 150 randomly distributed sine and square waveforms with a period of $T=1.2$ s, which was divided into training and testing subsets. During training, the filling factor response to the current waveform was sampled (green data points in the top panel of Fig.\ \ref{fig7}(a)), and weights (\textit{w}$_\text{C}$ and \textit{w}$_\text{P}$) for the current and previous waveforms were calculated using the Moore-Penrose pseudo-inverse matrix \cite{Sun2023}(see Appendix A: Methods). These weights were then used to classify the test dataset. The recognition accuracy was determined by comparing the predicted waveforms with the actual input waveforms. 

Figure\ \ref{fig7}(b) summarizes the recognition accuracy for various voltage ranges and magnetic bias fields. For the current waveform, the accuracy consistently exceeds 80\%, whereas for the previous waveform, it falls to $50\%-70\%$. To explore the effect of voltage range on device performance, Fig.\ \ref{fig7}(c) compares the filling factor evolution for two voltage ranges, 0 to 2 V and -0.5 to 1.5 V, under a 0.8 mT magnetic field. The output signal is more pronounced and less noisy in the -0.5 to 1.5 V range, enabling clear differentiation between sine and square waveforms. This voltage range dependence affects the recognition accuracy differently for current and previous waveforms. Comparing the 0 to 2 V and -0.5 to 1.5 V ranges at 0.8 mT, the -0.5 to 1.5 V range achieves better recognition accuracy for the current waveform (99\% versus 97\% ), while the 0 to 2 V range performs better for the previous waveform (66\% versus 57\%). This behavior reflects the trade-off between memory decay and signal robustness. A lower peak voltage (1.5 V) accelerates memory decay (as shown in Fig.\ \ref{fig4}(c)), reducing interference from prior states and enhancing current waveform recognition. However, stronger memory retention at higher voltages (2 V) improves recognition of the previous waveform by mitigating overwriting effects. 

To further assess the influence of voltage range on waveform classification, we compare the results for three voltage ranges centered around 0.5 V but with different peak-to-peak amplitudes: -0.5 to 1.5 V, -0.25 to 1.25 V, and 0 to 1.0 V (Fig.\ \ref{fig7}(b)). These ranges correspond to peak-to-peak amplitudes of 2 V, 1.5 V, and 1 V, respectively. Across all ranges, a magnetic field of 0.6 mT yields the highest recognition accuracy for the current waveform. At this field, voltage cycling generates a mixed magnetic state in the CoFeB layer, comprising stripe domains and skyrmions, which optimally transform the input signal for classification. For peak-to-peak amplitudes of 1.5 V and 2 V, the recognition accuracy for the current waveform reaches 100\%, indicating that these voltage ranges induce sufficiently distinct changes in the filling factor to ensure reliable classification. In contrast, classification results for the previous waveform exhibit greater variation and are generally less accurate. This reduced accuracy is particularly pronounced when the peak-to-peak voltage modulation in the input is too small to elicit a sufficiently large response in the filling factor. In such cases, the overlap between the magnetic states corresponding to different prior waveforms increases, making it more challenging to reliably differentiate between them. This observation highlights the sensitivity of the device's memory dynamics to input signal strength and underscores the need to optimize voltage modulation for tasks requiring high accuracy in previous waveform classification.  

To further enhance classification performance, we explore the use of multiple magneto-ionic devices operating under different magnetic bias fields, as suggested in previous studies \cite{Yokouchi2022}. This approach is emulated by combining data from experiments conducted at 0.4 mT, 0.6 mT, and 0.8 mT to create a single data set for each voltage range on which the training and recognition tasks are carried out. As shown in Fig.\ \ref{fig7}(d), this multi-field strategy achieves nearly 100\% recognition accuracy for the current waveform across all tested voltage ranges, representing a substantial improvement over single-field operation (Fig.\ \ref{fig7}(b)). The most notable enhancement is observed in the 0 to 1 V range, where accuracy increases from 84-92\% in single-field operation to a perfect 100\% when three devices operating under different fields are combined. An alternative strategy involves running multiple devices at different voltage ranges, which is emulated here by combining the data taken at different voltage ranges. As illustrated in Fig.\ \ref{fig7}(e), this approach also significantly improves the classification performance. However, recognizing previous waveforms remains challenging under both strategies, with accuracy consistently limited to the $50-60$\% range. These findings indicate that improving the recognition of previous waveforms may require innovative approaches beyond multi-device operation.  

To achieve substantial improvements in waveform classification, particularly for previous waveforms, increasing the number of sampling points per waveform is crucial. However, the 16 Hz frame-capturing rate of MOKE microscopy necessitates extending the waveform period beyond 1.2 s to accommodate more sampling points. While this extension is theoretically viable, it introduces practical challenges such as image drift and compromises memory fidelity due to intrinsic memory fading on a timescale of a few to a few tens of seconds (Fig.\ \ref{fig4}(c, f)). A more promising strategy involves shortening the waveform period to enable faster signal recognition while preserving intrinsic memory. This approach demands a readout mechanism with significantly higher temporal resolution than MOKE microscopy. Techniques such as electrical detection via the anomalous Hall effect, as demonstrated in prior studies \cite{Yokouchi2022,Sun2023}, or tunneling magnetoresistance, present compelling alternatives. These methods not only facilitate high-speed signal processing but also offer compatibility with large-scale device integration, paving the way for practical neuromorphic computing applications. 

\section{Summary}
We introduced a magneto-ionic supercapacitor structure with properties that are suited for applications in synaptic devices and reservoir computing. Its functionality is based on the nucleation and annihilation dynamics of stripe domains and magnetic skyrmions in a perpendicularly magnetized CoFeB. These processes are governed by reversible voltage-driven migration of Li\(^+\) ions, which alters the perpendicular magnetic anisotropy, and the strength of an applied magnetic bias field. The interplay between voltage, magnetic field, and the unique energetics of stripe domains and skyrmions creates a versatile platform for fine-tuning the system's nonlinearity and memory capacity. Exploiting these capabilities, we demonstrated both short-term and long-term adjustments of synaptic weights, as well as the classification of voltage waveforms in a reservoir computing tasks. While the results presented in this work were obtained using magneto-optic imaging, the device's crossbar design enables the incorporation of electrical readout mechanisms. This advancement would not only accelerate operation but also pave the way for integration into large-scale, energy-efficient neuromorphic networks.

\section*{Supplementary Material}
The supplementary material includes cyclic voltammetry measurements of the device, MOKE images showing the magnetic state under various voltage inputs and magnetic bias fields, and details on extracting the signal decay time constant following voltage pulsing.

\begin{acknowledgments}
This project has received funding from the European Union's Horizon 2020 research and innovation program under the Marie Sk\l{}odowska-Curie grant agreement No. 860060 “Magnetism and the effects of Electric Field” (MagnEFi) and was supported by the Research Council of Finland (Grant No. 338748). We acknowledge the provision of facilities by Aalto University at OtaNano - Micronova Nanofabrication Centre.
\end{acknowledgments}

\section*{Data Availability}
The data that support the findings of this study are available from the corresponding author upon reasonable request. 

\appendix

\section{Methods}
\subsection{Fabrication}
Multilayer stacks of Ta (5 nm) / Co\(_{20}\)Fe\(_{60}\)B\(_{20}\) (0.96 nm) / Ta (0.08 nm) / MgO (2 nm) / Ta (2 nm) were deposited on Si/SiO\(_2\) substrates via magnetron sputtering in a Singulus Rotaris system. The multilayer film was patterned into bottom electrodes using photolithography and ion beam milling. Top electrodes consisting of LiPON (70 nm) / Pt (5 nm) were fabricated atop the bottom electrodes, forming crossbar junctions. This process involved photolithography, magnetron sputtering in a Kurt J.\ Lesker system, and a lift-off technique. All layers were grown at room temperature in an argon atmosphere, except for the LiPON layer, which was deposited using RF sputtering from a Li$_3$PO$_4$ target with N\(_2\) as the process gas.

\subsection{Electrical and magneto-optical characterization}
The ionic properties of the crossbar junctions were analyzed using cyclic voltammetry performed with a Keithley 2450 sourcemeter. Data presented in \textcolor{blue}{Supplementary Fig. S1} were collected after extensive voltage cycling. Voltages for cycling, pulsing, and waveform classification experiments were generated using a Digilent Analog Discovery 2 instrument. The magnetic states of the CoFeB layer were visualized with an Evico magneto-optic Kerr effect (MOKE) microscope, equipped with a white-light source and a 100$\times$ objective lens. A 1.6$\times$ projection lens was used before the camera. A background image was captured before each video and subtracted to provide a magnetic contrast image. MOKE images were captured as video sequences at a frame rate of $12-16$ frames per second.

\subsection{MOKE image processing}
MOKE images were processed in Python using libraries including NumPy, scikit-image (SKIMAGE), OpenCV, PIMS, and SciPy. The raw images were first cropped. Long-term drift in the MOKE imaging was mitigated by identifying surface defects and replacing them with the mean grayscale value. Contrast limited adaptive histogram equalization (CLAHE) was then applied to enhance local contrast, followed by Gaussian blurring to reduce noise. A binary mask was created through thresholding, and small holes in the mask were removed using the \texttt{skimage.morphology.remove\_small\_holes} function, which fills holes smaller than a specified size to enhance the integrity of detected features. The filling factor (FF) was subsequently calculated from the binary mask using the formula:

\[
    FF = \mathrm{\frac{Number\ of\ feature\ pixels}{Total\ number\ of\ pixels} }
\]
To improve image quality and minimize noise in the calculated filling factor, the data presented in Fig.\ \ref{fig4} was derived by averaging five consecutive MOKE images. Specifically, for each frame \textit{i}, pixel values from frames indexed \textit{i}-2 to \textit{i}+2 were averaged before further processing.

\subsection{Waveform recognition task}
The voltage input for the waveform classification task (Fig.\ \ref{fig7}) consisted of 150 randomly ordered sine and square waveforms, each with a period of 1.2 s. The device's output signal, represented by the filling factor derived from MOKE imaging, was utilized for classification. To preserve the highest possible temporal resolution, MOKE frames were not averaged in this experiment. The classification objective was to distinguish between sine and square waveforms, assigned binary indicator values of 1 and 0, respectively. The output signal for the current waveform was sampled at the MOKE frame-capturing rate, and weights \textit{w}$_{C}$ and \textit{w}$_{P}$ were calculated for the classification of current and previous waveforms, respectively. The weights were obtained by solving the linear Moore-Penrose pseudo-inverse matrix using the following relation:
\[
    \begin{bmatrix} 
    w_{C}^T \\ 
    \\
    w_{P}^T 
    \end{bmatrix}
    \begin{bmatrix}
    s_1\\
    s_2\\
    \vdots\\
    s_n
    \end{bmatrix}
    =
    \begin{bmatrix} 
    C \\ 
    \\
    P 
    \end{bmatrix}
\]

\noindent where $s_1$, $s_1$, ... $s_n$ represent the sampling points (captured evolution of the filling factor), and $C$ and $P$ denote the binary indicator values for the current and previous waveforms. Once the weight matrix was determined during training, it was applied to classify the waveforms in the test dataset. Recognition accuracy was calculated as:
\[
    \text{Accuracy (\%) }= \frac{\text{Number\ of\ right\ predictions}}{\text{Size\ of\ the\ test\ dataset}} \times 100 \%
\]

\noindent To ensure robust performance evaluation, the dataset was randomized, and five-fold cross-validation was conducted. The data was split into five equal parts, with four used for training and one for testing in each fold. The average result across five folds were used to determine the recognition accuracy, and the error bars in Fig.\ \ref{fig7} were derived from the variation across these folds.

\bibliography{paper3}
\end{document}